\documentstyle[12pt,sidefig,psfig,epsf]{article}

\catcode`\@=11
\long\def\@makefntext#1{
\protect\noindent \hbox to 3.2pt {\hskip-.9pt

$^{{\ninerm\@thefnmark}}$\hfil}#1\hfill}                

\def\@makefnmark{\hbox to 0pt{$^{\@thefnmark}$\hss}}  

\def\ps@myheadings{\let\@mkboth\@gobbletwo
\def\@oddhead{\hbox{}
\rightmark\hfil\ninerm\thepage}
\def\@oddfoot{}\def\@evenhead{\ninerm\thepage\hfil
\leftmark\hbox{}}\def\@evenfoot{}
\def\sectionmark##1{}\def\subsectionmark##1{}}

\renewcommand{\thefootnote}{\fnsymbol{footnote}}

\newcounter{sectionc}\newcounter{subsectionc}\newcounter{subsubsectionc}

\renewcommand{\section}[1] {\vspace*{0.6cm}\addtocounter{sectionc}{1}
\setcounter{subsectionc}{0}\setcounter{subsubsectionc}{0}\noindent
        {\normalsize\bf\thesectionc. #1}\par\vspace*{0.4cm}}

\renewcommand{\subsection}[1] {\vspace*{0.6cm}\addtocounter{subsectionc}{1}
        \setcounter{subsubsectionc}{0}\noindent
        {\normalsize\it\thesectionc.\thesubsectionc. #1}\par\vspace*{0.4cm}}

\renewcommand{\subsubsection}[1]
    {\vspace*{0.6cm}\addtocounter{subsubsectionc}{1}
     \noindent {\normalsize\rm\thesectionc.\thesubsectionc.\thesubsubsectionc.
        #1}\par\vspace*{0.4cm}}

\newcounter{appendixc}

\newcounter{subappendixc}[appendixc]

\newcounter{subsubappendixc}[subappendixc]

\renewcommand{\appendix}[1] {\vspace*{0.6cm}
        \refstepcounter{appendixc}
        \setcounter{figure}{0}
        \setcounter{table}{0}
        \setcounter{equation}{0}
        \renewcommand{\thefigure}{\Alph{appendixc}.\arabic{figure}}
        \renewcommand{\thetable}{\Alph{appendixc}.\arabic{table}}
        \renewcommand{\theappendixc}{\Alph{appendixc}}
        \renewcommand{\theequation}{\Alph{appendixc}.\arabic{equation}}
        \noindent{\bf Appendix \theappendixc #1}\par\vspace*{0.4cm}}

\def\abstracts#1{{
        \centering{\begin{minipage}{12.2truecm}\footnotesize\baselineskip=12pt\noindent
        \parindent=0pt #1
        \end{minipage}}\par}}

\renewenvironment{thebibliography}[1]
       {\begin{list}{\arabic{enumi}.}
        {\usecounter{enumi}\setlength{\parsep}{0pt}
\setlength{\leftmargin 1.25cm}{\rightmargin 0pt}
         \setlength{\itemsep}{0pt} \settowidth
        {\labelwidth}{#1.}\sloppy}}{\end{list}}

\newcounter{itemlistc}
\newcounter{romanlistc}
\newcounter{alphlistc}
\newcounter{arabiclistc}

\newcommand{\fcaption}[1]{
        \refstepcounter{figure}
        \setbox\@tempboxa = \hbox{\footnotesize Fig.~\thefigure. #1}
        \ifdim \wd\@tempboxa > 6in
           {\begin{center}
        \parbox{6in}{\footnotesize\baselineskip=12pt Fig.~\thefigure. #1}
            \end{center}}
        \else
             {\begin{center}
             {\footnotesize Fig.~\thefigure. #1}
              \end{center}}
        \fi}

\newcommand{\tcaption}[1]{
        \refstepcounter{table}
        \setbox\@tempboxa = \hbox{\footnotesize Table~\thetable. #1}
        \ifdim \wd\@tempboxa > 6in
           {\begin{center}
        \parbox{6in}{\footnotesize\baselineskip=12pt Table~\thetable. #1}
            \end{center}}
        \else
             {\begin{center}
             {\footnotesize Table~\thetable. #1}
              \end{center}}
        \fi}

\def\@citex[#1]#2{\if@filesw\immediate\write\@auxout
        {\string\citation{#2}}\fi
\def\@citea{}\@cite{\@for\@citeb:=#2\do
        {\@citea\def\@citea{,}\@ifundefined
        {b@\@citeb}{{\bf ?}\@warning
        {Citation `\@citeb' on page \thepage \space undefined}}
        {\csname b@\@citeb\endcsname}}}{#1}}

\newif\if@cghi
\def\cite{\@cghitrue\@ifnextchar [{\@tempswatrue
        \@citex}{\@tempswafalse\@citex[]}}
\def\citelow{\@cghifalse\@ifnextchar [{\@tempswatrue
        \@citex}{\@tempswafalse\@citex[]}}
\def\@cite#1#2{{$\null^{#1}$\if@tempswa\typeout
        {IJCGA warning: optional citation argument
        ignored: `#2'} \fi}}

 1
 1
 1

\font\ninerm=cmr9

\begin{document}

\centerline{\large\bf 
Ab initio shallow acceptor levels  in gallium nitride}

\baselineskip=22pt

\baselineskip=16pt
\vspace*{0.18cm}
\centerline{\normalsize  Vincenzo Fiorentini,$^1$  Fabio Bernardini,$^1$ 
 Andrea Bosin,$^2$ and David Vanderbilt$^3$}
\vspace*{0.01cm}
\baselineskip=13pt
\begin{center}
{\footnotesize\it (1) Istituto Nazionale di Fisica della Materia --
 Dip.  Scienze Fisiche,
Universit\`a di Cagliari, Italy}\\
\baselineskip=12pt
{\footnotesize\it (2) 
 Istituto Nazionale di Fisica della Materia  --
SISSA, Trieste, Italy} \\
{\footnotesize\it (3) 
 Department of Physics and Astronomy, Rutgers University, 
Piscataway, NJ, U.S.A.}
\end{center}

\vspace*{0.2cm}

\abstracts{Impurity
 levels and formation energies of acceptors  in wurtzite GaN are 
predicted ab initio.  Be$_{\rm Ga}$ is found to be the
shallow (thermal ionization energy
$\sim$ 0.06 eV); 
 Mg$_{\rm Ga}$ and  Zn$_{\rm Ga}$ are 
 mid-deep acceptors (0.23 eV and 0.33 eV respectively);
 Ca$_{\rm Ga}$ and Cd$_{\rm Ga}$  are deep 
acceptors  ($\sim$0.65 eV); 
Si$_{\rm N}$ is a midgap trap with high formation energy;
finally, contrary to recent claims, C$_{\rm N}$ is a deep acceptor
(0.65 eV).
Interstitials and heteroantisites  are energetically
not  competitive with substitutional incorporation.
}

\normalsize\baselineskip=15pt
\setcounter{footnote}{0}
\renewcommand{\thefootnote}{\alph{footnote}}

\vspace{-0.3cm}
\section{Introduction}
\vspace{-0.3cm}

 Gallium nitride has recently emerged as a  serious
candidate for the fabrication of green to violet light 
emitting devices. Blue to yellow-green LED's\cite{nakaled,nakalas} and, 
most recently, a GaN-based blue laser have been 
demonstrated.\cite{nakalas}
A basic problem that needs to be solved
 is that of reproducible and 
efficient $p$-type doping of GaN.
Mg was succesfully
employed as $p$-dopant  for device applications\cite{nakamg}
only after elaborate post-growth treatments,
and despite its large ($\sim$0.25 eV)   ionization energy. The
relevant basic  mechanisms are  still under scrutiny.\cite{neuapl}

There is thus an  obvious interest in shallower acceptors for
 GaN, and first-principle theoretical 
investigations can reliably indicate suitable candidates
to this end.
 In this paper we present a preliminary account of
ab initio studies on  impurities from
Groups IIA, IIB, and IV in wurtzite GaN. 
Our main results are: Be$_{\rm Ga}$,
with a thermal level at 0.06 eV  above the valence band,
is the shallowest acceptor reported so far in GaN; 
Mg$_{\rm Ga}$ and Zn$_{\rm Ga}$ are found to be
mid-deep acceptors (0.23 eV and 0.33 eV respectively)
in good agreement with experiments;
C$_{\rm N}$, Ca$_{\rm Ga}$, and Cd$_{\rm Ga}$  are deep 
acceptors  (around 0.65 eV);
Si$_{\rm N}$ is a midgap trap (1.97 eV) with high formation energy.
Importantly, preliminary results on interstitials
and heteroantisites  indicate that these are energetically
not  competitive with substitutional incorporation.

\vspace{-0.4cm}\section{Method}\vspace{-0.3cm}

The calculation of acceptor levels and formation energies is 
non-trivial, especially in a computationally demanding system such as
GaN.  We  calculate local-density-functional total energies and
forces with a conjugate-gradient 
minimization scheme,\cite{ksv} plane-waves,
and  ultrasoft pseudopotentials.\cite{vand} 
By the latter, we  describe  accurately all ``difficult''
electronic states (such as Ga and Zn 3$d$,  N 2$p$, Be 2$s$,
Cd 4$d$, Mg and Ca semicore $s$ and $p$, etc.) with a cutoff of 
25 Ryd. The defects are simulated by single impurities in
  32-atom wurtzite GaN supercells.
Full geometry optimization  is performed in all cases,
 using a newly developed method (details given elsewhere).
The impurity formation energy in the charge state {\it Q}
 ($Q$ is the
number of electrons transferred {\it from} the defect {\it to} a
reservoir of chemical potential $\mu_e$)  is given by 
\begin{equation}
E_f(Q) = 
E^{\rm tot}(Q)  -n^{\rm Ga} \mu^{\rm Ga}   -n^{\rm N} \mu^{\rm N}-n^{\rm X} 
\mu^{\rm X} + Q (\mu_e + E_v),
\label{eform}
\end{equation}
where
$E^{\rm tot}(Q)$ is the defected supercell total energy in the specific
 charge state, $\mu_e$ the electron chemical potential, 
$E_v$ the top valence band energy,
$n^{\rm Ga}$,  $n^{\rm N}$, and $n^{\rm X}$
are the number of Ga,  N, and X atoms, and $\mu^{\rm Ga}$, $\mu^{\rm N}$, 
and $\mu^{\rm X}$ their chemical potentials.
These potentials  must satisfy the 
equilibrium conditions with GaN and with appropriate solubility-limiting  
X-N or X-Ga compounds (whose properties are also calculated ab
initio). To study the case of maximum impurity solubility, 
we fix  $\mu_{\rm X}$  at the maximum value
compatible with the solubility limit; in this case
a single free potential remains (e.g. $\mu_{\rm N}$), which is determined 
by the imposed growth conditions.

The thermal ionization
level $\epsilon(0/-)$
of a   single acceptor
is by definition the formation energy difference of the two 
charge states $Q$=0 and $Q$=--1 
at $\mu_e$=0, or otherwise stated the electron chemical potential 
value at  which the negatively charged state becomes favored over 
the neutral one. This effectively corresponds
to the capture of an electron in the center with concurrent
 geometric relaxation. The
 optical level, accessible  in
luminescence experiments, is obtained summing the calculated 
Franck-Condon shift to the thermal energy.
To refer the energies of different charge states
to the same zero, we align the supercell
 local potentials  in appropriate bulk-like regions.
 By  bulk-like region we mean one where the macroscopic 
average\cite{bald} of the local potential is flat and 
unaffected by the impurity. Such  regions are chosen
on the wurtzite $c$-axis, along which
the  impurity state decays faster (see below),
and the inter-acceptor distance 
is  maximal  ($\sim$ 20 bohr).
  Fig.\,1 exemplifies the procedure for Zn: a  
plateau appears  halfway between the impurities (placed at 0 and 1), 
and it  is there that we pick the potential values for the alignment.

\begin{sidefigure}
\unitlength=1cm
\vspace{-2.5cm}\hspace{-0.6cm}
\begin{picture}(7.5,7.5)
\put(0.0,0.2){\psfig{figure=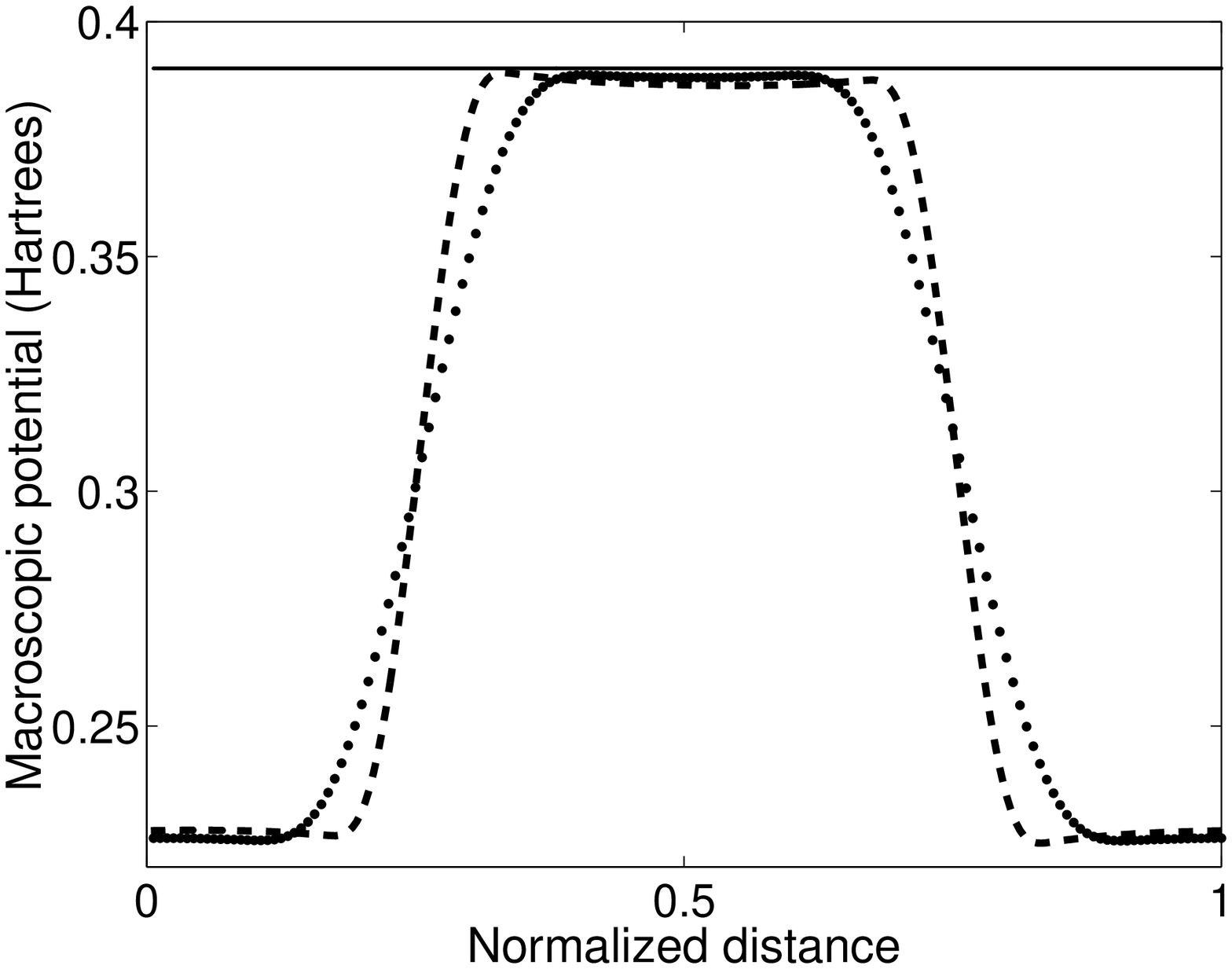,width=6.6cm,height=4.6cm}}
\end{picture}
\parbox{6.5cm}{\footnotesize\baselineskip=12pt Fig.~1:
~Local potential macroscopic averages
on $c$-axis for bulk GaN (solid) and Zn$^0$ (dashed), and 
on
 $a$-axis for Zn$^{0}$ (dotted).}
\end{sidefigure}
Ideally, one should reach    bulk-like regions  
 with the {\it  same} potential value
no matter in which direction one moves away from the 
impurity;  an upper bound  of 
the alignment uncertainty is thus the difference of the constant 
 values along e.g. the $a$ and $c$ axis.
This directional  deviation 
is essentially all of our ``formal'' error bar 
in the calculated levels, as our calculations are 
otherwise well converged in terms of technical ingredients.
It amounts  to $\sim$ 0.05 eV for Zn (Fig.\,1), and is generally
of order 0.1 to 0.2 eV at worst.

\vspace{-0.4cm}\section{Results and Discussion}\vspace{-0.3cm}

In Table 1, we summarize the results
 for the acceptors of groups II-A, II-B, and IV.
A new candidate $p$-dopant emerges:  Be$_{\rm Ga}$, with a
thermal level at $E_v$ + 0.06 eV. Mg and Zn are progressively deeper
(0.23 eV and 0.33 eV), while all others are definitely deep.
In $n$-type conditions ($\mu_e\sim E_{\rm gap} \simeq 3.4$ eV)
 the formation energies of all acceptors in their negative charge state
(except Si$_{\rm N}$) are 
low or negative, which makes them efficient donor 
compensators. While in $p$-type and Ga-rich conditions  compensation by
nitrogen vacancy donors
 is known to be a problem,
{\small
\begin{table}[tbh]
\centering
\tcaption{Acceptors in wurtzite GaN.  
Formation energies E$_{f}$  of 
the neutral charge state in N-rich conditions,
thermal impurity levels, relaxation contribution E$_{r}$, change in 
distance to neighbors in the {\it a}-plane ($\Delta$d$_{a}$) and 
along the {\it c}-axis 
($\Delta$d$_{c}$) in percentage of the ideal bond length,
solubility limits and their formation enthalpy. All energies are in eV.}
\vspace{0.15cm}
\small
\begin{tabular}{|l|r|r|r|r|r|l|r|} \hline\hline
Defect & E$_f$ & $\epsilon$(0/--) & E$_r$  & $\Delta$ d$_{a}$ & $\Delta$ 
d$_{c}$ & limit & $-\Delta$H$_f$\\ \hline 
Be$_{\rm Ga}$    & 2.29 & 0.06 & 0.40 & ---5.5 & --8.4 
 & Be$_3$N$_2$ & 6.70 \\ 
Mg$_{\rm Ga}$    & 1.40 & 0.23 & 0.21 & +3.2 & +3.2  & Mg$_3$N$_2$ & 4.96 \\ 
Zn$_{\rm Ga}$    & 1.21 & 0.33 & 0.03 & +1.1 & --0.5 & Zn$_3$N$_2$ & 1.28 \\
Ca$_{\rm Ga}$    & 2.15 & 0.62 & 1.87 & +10.0& +12.7 & Ca$_3$N$_2$ & 4.91 \\
Cd$_{\rm Ga}$    & 1.60 & 0.65 & 1.40 & +9.2 & +8.8  & h-Cd    &     \\
\hline
C$_{\rm N }$     & 4.24 & 0.65 & 0.04 & +0.5 & --0.6  & d-C    &     \\
Si$_{\rm N }$    & 5.49 & 1.97 & 3.36 &+12.0 & +11.8 & d-Si   &     \\
\hline
V$^{+}_{\rm N}$ & 2.32 &      & 0.07 &  0.0 &  +1.2 &   w-GaN        & 1.65  \\
\hline\hline
\end{tabular}
\end{table}}
from calculations of  impurity concentrations accounting for 
the formation of singly-positive  nitrogen vacancies
(formation energies: 2.32 eV (N-rich) and 0.67 eV (Ga-rich) at $\mu_e=0$)
and for  overall neutrality,\cite{znse}
we find that acceptor-vacancy compensation in $p$-type conditions
is essentially suppressed  for all species in N-rich conditions.
In $p$-type conditions,  formation energies are 
 low for Zn and Mg, somewhat higher for Be,
high for C$_{\rm N}$, very high for Si$_{\rm N}$.
The carrier concentration at room temperature 
for the shallowest acceptor (Be)  
in N-rich conditions and at MOCVD growth temperatures,
ranges from   $\sim$3$\times$10$^{16}$ cm$^{-3}$
to  $\sim$4$\times$10$^{18}$ cm$^{-3}$
for a  formation entropy of 0 to 10 k$_{\rm B}$.
 The incorporation mechanism suggested\cite{neuapl}  for Mg may
be helpful also in the case of Be: the calculated formation energy 
of the Be-H complex (limits: H$_2$ and Be$_3$N$_2$) is as low as 
0.32 (0.87) eV/pair in N-rich (Ga-rich) conditions. The 
concentration of Be thus 
incorporated would grow by 3 orders of magnitude. Of course H
is also incorporated, and must 
be  detached from the complex to activate 
the acceptor: this 
requires about 1.5 eV  for the Be-H complex, so that  LEEBI or 
thermal  annealing may be needed.

 Contrary to recent claims\cite{haller}  on
the  C$_{\rm N}$ acceptor being shallow ($\sim$ 0.2 eV),
we find it to be deep, with a thermal
 ionization energy of 0.65 eV. Due to its high formation
energy in $p$-type conditions, any appreciable role of 
C$_{\rm N}$ as a shallow dopant, and in $p$-type conditions, 
is to be ruled out. On the other hand 
 its formation energy drops to near zero in $n$-type conditions,
so it   may be present in high concentrations in $n$--GaN.
 The  optically active level is  0.8 eV 
above the valence band.

The electronic structure of the acceptors is unusual 
in several respects. {\it First},  the 
impurity states of the Ga-substituting species
(including the mid-deep and shallow ones)  are largely localized on the
first-neighbor N shell.
{\it  Second}, the impurity states are
 of N $p_{xy}$--like character.
This is  a reminder of the impurity state's origin from 
the $\Gamma_6$ top valence band doublet;
as such, it is of course peculiar of wurtzite GaN.
The state is generally localized in 
the $a$-plane, and decays  rapidly along the $c$-axis.
We display it for Mg in Fig.\,2.
The unusual shape of the impurity state affects
the properties of H-acceptor complexes: e.g.,
H's preferential binding at the N-an\-ti\-bon\-ding site in the $a$-plane 
rather than along  the $c$-axis, 
correlates with the $a$-plane localization  of the state accepting 
the extra electron. {\it Third},
in all cases, we find a sizable polarization of the 
3$d$ states of second-neighbor Ga's.
For  Zn, the 3$d$  states are even shallower, and 
they mix into the impurity state. This  may 
be the cause of the larger Zn binding energy in
GaN as compared to e.g. GaAs. In the latter,
 the delocalized carrier bound to the shallow center
 sees the Ga and Zn cores
as effectively identical:
this may not hold in GaN, where the  3$d$'s
hybridize with $sp$ valence states.

\begin{sidefigure}
\unitlength=1cm
\vspace{-2.5cm}\hspace{-0.6cm}
\begin{picture}(7.5,7.5)
\put(0.0,0.2){\psfig{figure=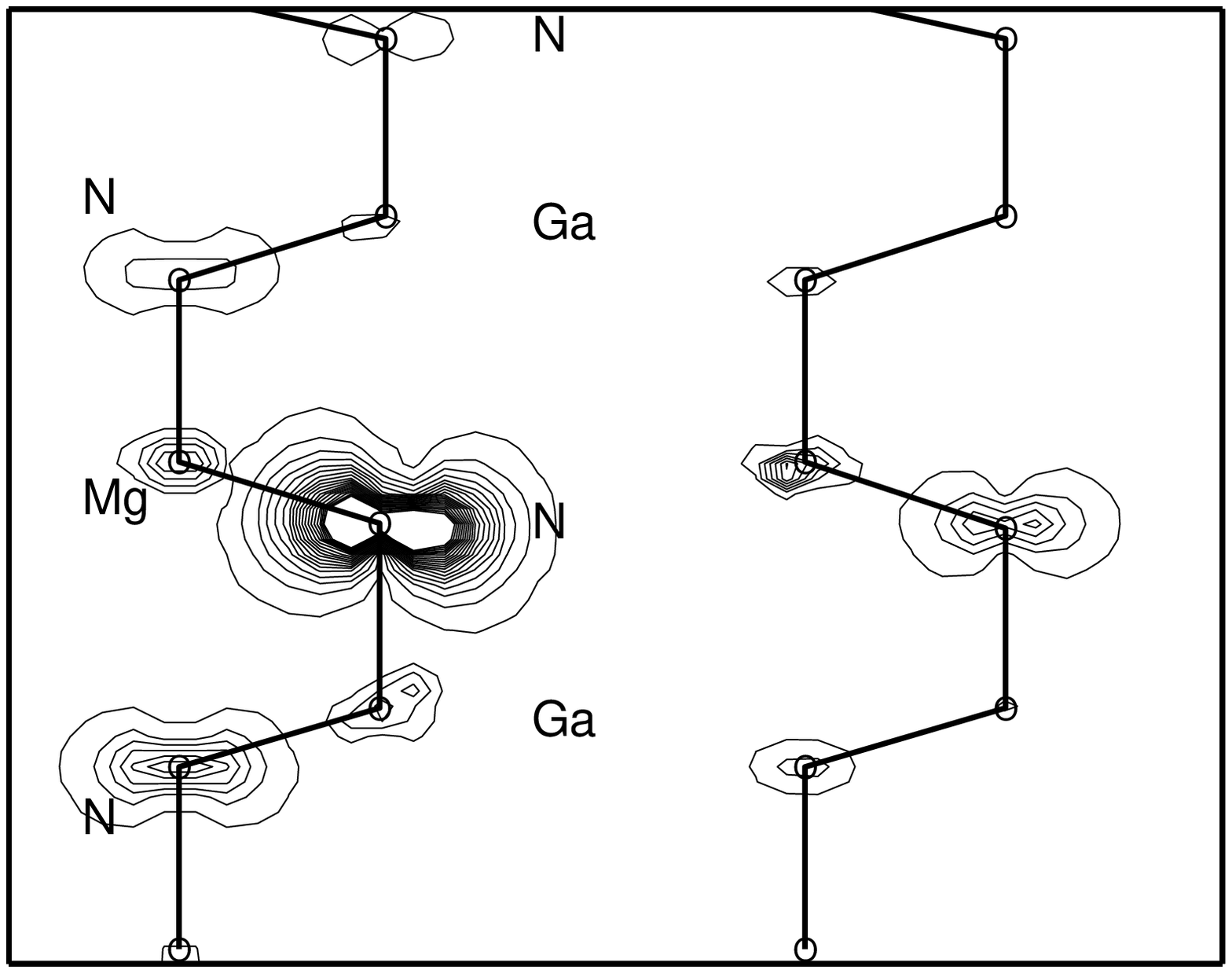,width=6.6cm,height=4.6cm}}
\end{picture}
\parbox{6.5cm}{\footnotesize\baselineskip=12pt Fig.~2:
~Number density of the impurity state for Mg. Note the $p_{xy}$-like
character and the localization on the first-neighbor shell.}
\end{sidefigure}
Finally, whether or not acceptors in GaN behave effective-mass like 
is an open issue. It may be speculated that the answer is 
 in the affirmative: using an  average
hole mass $\overline{{\rm m}}_h\sim 0.8$ m$_e$ 
the hydrogenic binding energy is 0.4 eV (not far from 
the ``isocoric'' acceptor Zn) when 
the high-frequency dielectric constant
$\varepsilon\!\equiv\!\varepsilon_{\infty}\!=\!5.4$ is used.
This is suggested by the fact that most acceptor binding energies 
are much  larger than typical GaN phonon energies.\cite{orton}
 As for the shallower Be center, Be is expected to have repulsive 
central cell corrections, and  its smaller binding energy may 
 imply using a low-frequency dielectric constant
$\varepsilon\!\equiv\!\varepsilon_{\circ}\!\sim\!10$ instead:
in this case the binding energy is in fact $\sim$ 0.1 eV. Of course
the valence band structure  and wave-vector--dependent screening  must be 
described in detail before any conclusion can be reached.

%

\end{document}